\documentclass[aps,pra,twocolumn,showpacs,superscriptaddress]{revtex4}

\usepackage{amsmath}
\usepackage{graphicx}

\begin{document}

\title{Unification of Dynamical Decoupling and the Quantum Zeno Effect}
\author{P. Facchi}
\affiliation{Dipartimento di Fisica, Universit\`a di Bari I-70126
  Bari, Italy, and Istituto Nazionale di Fisica Nucleare, Sezione di
  Bari, I-70126 Bari, Italy}
\author{D.A. Lidar}
\affiliation{Chemical Physics Theory Group, Chemistry Department,
  University of Toronto, 80 St. George Street, Toronto, Ontario M5S 3H6, Canada}
\author{S. Pascazio}
\affiliation{Dipartimento di Fisica, Universit\`a di Bari I-70126
  Bari, Italy, and Istituto Nazionale di Fisica Nucleare, Sezione di
  Bari, I-70126 Bari, Italy}
\date{\today}

\begin{abstract}
We unify the quantum Zeno effect (QZE) and the ``bang-bang" (BB)
decoupling method for suppressing decoherence in open quantum
systems: in both cases strong coupling to an external system or
apparatus induces a dynamical superselection rule that partitions
the open system's Hilbert space into quantum Zeno subspaces. Our
unification makes use of von Neumann's ergodic theorem and avoids
making any of the symmetry assumptions usually made in discussions
of BB. Thus we are able to generalize BB to arbitrary fast and
strong pulse sequences, requiring no symmetry, and to show the
existence of two alternatives to pulsed BB: continuous decoupling,
and pulsed measurements. Our unified treatment enables us to
derive limits on the efficacy of the BB method: we explicitly show
that the inverse QZE implies that BB can in some cases accelerate,
rather than inhibit, decoherence.
\end{abstract}

\maketitle

\newcommand{\andy}[1]{ }

\renewcommand{\Im}{\mathrm{Im}}

\section{Introduction}

Recent years have witnessed a surge of interest in ways to protect
quantum coherence, driven mostly by developments in the theory of
quantum information processing \cite{Nielsen:book}. A number of
promising strategies for combatting decoherence have been
conceived and in some cases experimentally tested, including
quantum error correcting codes and topological codes (for a review
see \cite{Preskill:99}), decoherence free subspaces and
(noiseless) subsystems (for a review see \cite{LidarWhaley:03}),
and ``bang-bang" (BB) decoupling
\cite{Viola:98,Vitali:99,Viola:99} (for an overview see
\cite{Viola:01a}). Two recent papers have shown that these various
methods can be unified under a general algebraic framework
\cite{Zanardi:03}. Here, using a very different approach, we
continue this development for BB decoupling and the quantum Zeno
effect (QZE).

The idea behind BB is that the application of sufficiently strong
and fast pulses, with appropriate symmetry (notions we make
precise later), when applied to a system, can decouple it from its
decohering environment. The notion of a strong and fast
interaction with a quantum system is also the key idea behind the
QZE \cite{Misra:77} (for reviews see \cite{HW,PIO}). The standard
view of the QZE effect is that by performing frequent projective
measurements one can freeze the evolution of a quantum state (``a
watched pot cannot boil''). However, recently it has become clear
that this view of the QZE is too narrow, in two main respects: (i)
The projective measurements can be replaced by another quantum
system interacting strongly with the principal system
\cite{Harris:82plSchulman98,PIO}; (ii) The states of the principal
system need not be frozen: instead the general situation is one of
dynamically generated quantum Zeno subspaces, in which non-trivial
coherent evolution can take place \cite{Facchi:PRL02}. It is
therefore not only physically reasonable, but also logically
appealing to view the QZE as a dynamical effect: in this broader
context, both BB decoupling and the QZE can be understood as
arising from the same physical considerations, and hence can be
unified under the same conceptual and formal framework.
Furthermore, they appear as particular cases of a more general
dynamics in which the system of interest is ``strongly" coupled to
an external system that (loosely speaking) plays the role of a
measuring apparatus.

We use these insights to (i) generalize the BB method to pulse
sequences with no symmetry; (ii) to point out that the BB pulses
can have the opposite from the desired effect (a situation well
known from the QZE literature as the ``inverse'' or ``anti'' Zeno
effect) \cite{IZE,IZEexp}; (iii) to show that alternatives to the
unitary pulse control scheme are available to suppress the
system-environment interaction, namely: a) \emph{continuous}
unitary interaction, and b) pulsed \emph{measurements}.

\section{Simplest BB cycle}

Consider the ``BB-evolution'' induced by
the two-element control set
(not necessarily a group) $\{I,U_{1}\}$, where $I$ is the identity
operator, in which the controlled system Q alternately
undergoes $N$ ``kicks" $U_{1}$ (\emph{instantaneous} unitary
transformations) and free evolutions in a time interval $t$
\begin{equation}
U_{N}(t)= [ U_{1}U(t/N) ] ^{N}.
\label{eq:BBevol}
\end{equation}
We take $U=\exp(-iHt)$, with $H$ the (time-independent)
Hamiltonian of Q, its environment and their interaction, and will
sometimes abbreviate $U(t/N) $ by $U$. We
present a new derivation of this ``BB-evolution'' that allows for
a transparent connection to the formulation of the QZE.

In the large $N$ limit, the dominant contribution to $U_{N}(t)$ is
$ U_{1}^{N} $. We therefore consider the sequence of unitary
operators
\begin{equation}
V_{N}(t)=U_{1}^{\dagger N}U_{N}(t).  \label{eq:sequence}
\end{equation}
Observe that $V_{N}(0)=I$ for any $N$ and
\begin{eqnarray}
i\frac{d}{dt}V_{N}(t) &=&U_{1}^{\dagger
N}\sum_{k=0}^{N-1}(U_{1}U)^{k}\left(
U_{1}i\frac{dU}{dt}\right) (U_{1}U)^{N-k-1}  \notag \\
&=&U_{1}^{\dagger
N}\frac{1}{N}\sum_{k=0}^{N-1}(U_{1}U)^{k}U_{1}HU_{1}^{
\dagger }(U_{1}U)^{\dagger k}(U_{1}U)^{N}  \notag \\
&=&H_{N}(t)V_{N}(t),\qquad (V_{N}(0)=I)
\end{eqnarray}
with
\begin{equation}
H_{N}(t)=\frac{1}{N}\sum_{k=0}^{N-1}U_{1}^{\dagger N}
(U_{1}U)^{k}U_{1}HU_{1}^{\dagger }(U_{1}U)^{\dagger k}U_{1}^{N}.
\label{eq:H_N}
\end{equation}
The limiting evolution operator
\begin{equation}
\mathcal{U}(t)\equiv \lim_{N\rightarrow \infty }V_{N}(t)
\label{eq:limseq}
\end{equation}
satisfies the equation
\begin{equation}
i\frac{d}{dt}\mathcal{U}(t)=H_{Z}\mathcal{U}(t),\qquad
(\mathcal{U}(0)=1)
\label{eq:eqUz}
\end{equation}
with the ``Zeno" Hamiltonian
\begin{equation}
H_{Z}\equiv \lim_{N\rightarrow \infty }H_{N}(t).
\end{equation}
Therefore $\mathcal{U}(t)=\exp (-iH_{Z}t)$. In order to study the
behavior of the limiting operator we first observe that for
$N\rightarrow \infty $ we can neglect the free evolution $U(t/N) $
in Eq.~(\ref{eq:H_N}) and so
\begin{equation}
H_{N}\sim \frac{1}{N}\sum_{k=0}^{N-1}U_{1}^{\dagger
N}U_{1}^{k+1}HU_{1}^{\dagger k+1}U_{1}^{N}=\frac{1}{N}
\sum_{k=0}^{N-1}U_{1}^{\dagger k}HU_{1}^{k}.
\end{equation}
Next we will show that for any bounded $H$ and any $U_1$ with a
pure point spectrum, namely
\begin{eqnarray}
U_{1}=\sum_{\mu }e^{-i\lambda _{\mu }}P_{\mu }
\label{eq:specdec}
\end{eqnarray}
[$\lambda _{\mu }\neq \lambda _{\nu }$ (mod $2\pi$) for $\mu \neq
\nu$, $P_{\mu }P_{\nu }=\delta _{\mu \nu }P_{\mu }$], one gets
\begin{equation}
H_{Z}=\lim_{N\rightarrow \infty
}\frac{1}{N}\sum_{k=0}^{N-1}U_{1}^{\dagger k}HU_{1}^{k}=\sum_{\mu
}P_{\mu }HP_{\mu }\equiv \Pi _{U_{1}}(H),
\label{eq:HZ}
\end{equation}
where the map $\Pi _{U_{1}}$ is the projection onto the
centralizer (or commutant) of $U_{1}$,
\begin{equation}
Z(U_{1})=\{X|\;[X,U_{1}]=0\} .
\end{equation}
First we show that the (strong)
limit $H_{Z}$ in Eq.~(\ref{eq:HZ}) is a bounded operator which
satisfies the intertwining property
\begin{equation}
H_{Z}P_{\mu }=P_{\mu }HP_{\mu }=P_{\mu }H_{Z}  \label{eq:intertwin}
\end{equation}
for any eigenprojection $P_{\mu }$ of $U_{1}$, with eigenvalue
$e^{-i\lambda _{\mu }}$. Equation (\ref{eq:HZ}) follows whenever
$U_{1}$ admits the spectral decomposition (\ref{eq:specdec}). Here
is the proof. For any vector $\psi$ in the Hilbert space
$\mathcal{H}$, we get, using Eq.~(\ref{eq:specdec})
\begin{equation}
\frac{1}{N}\sum_{k=0}^{N-1}U_{1}^{\dagger k}HU_{1}^{k}P_{\mu
}\psi = \frac{1}{N}\sum_{k=0}^{N-1}\tilde{U}^{k}\phi ,
\label{eq:proof}
\end{equation}
where $\tilde{U}=(U_{1}e^{i\lambda _{\mu }})^{\dagger }$ is a
unitary operator whose eigenprojection $P_{\mu }$ has eigenvalue
$1$ and $\phi =HP_{\mu }\psi \in \mathcal{H}$. Recall now an
ergodic theorem due to von Neumann \cite[p.~57]{ReedSimon} that
states that if $\tilde{U}$ is a unitary operator on the Hilbert
space $\mathcal{H}$ and $P_{\mu }$ its eigenprojection with
eigenvalue $1$ ($\tilde{U}P_{\mu }=P_{\mu }$), then for any $\phi
\in \mathcal{H}$
\begin{equation}
\lim_{N\rightarrow \infty
}\frac{1}{N}\sum_{k=0}^{N-1}\tilde{U}^{k}\phi =P_{\mu }\phi .
\label{eq:vNerg}
\end{equation}
As a consequence, by taking the limit of (\ref{eq:proof}), we get
(\ref{eq:intertwin}).

Notice that the intertwining property (\ref{eq:intertwin}) holds
also for an unbounded $H$ whose domain $D$ contains the range of
$P_\mu$, namely $P_{\mu}{\mathcal{H}}\subset D(H)$. For a generic
unbounded Hamiltonian, we can still formally consider
(\ref{eq:HZ}) as the limiting evolution, but the meaning of $P_\mu
H P_\mu$ and its domain of selfadjointness should be properly
analyzed.

In conclusion
\begin{equation}
\mathcal{U}(t)=\exp (-iH_{Z}t)=\exp [-i\sum_{\mu }P_{\mu }HP_{\mu
}t ]
\end{equation}
and, due to Eqs.~(\ref{eq:sequence}) and (\ref{eq:limseq}),
\begin{eqnarray}
U_{N}(t) &\sim& U_{1}^{N}\mathcal{U}=U_{1}^{N}\exp (-iH_{Z}t)
\nonumber \\
& =& \exp [ -i\sum_{\mu }(N\lambda _{\mu }P_{\mu }+P_{\mu
}HP_{\mu }t) ] .
\end{eqnarray}
This proves that the ``BB-evolution'' (\ref{eq:BBevol}) yields a
Zeno effect and a partitioning of the Hilbert space into ``Zeno
subspaces'', in the sense of \cite{Facchi:PRL02}.

We emphasize that no cyclic group properties are required for
pulse sequences. This extends previous studies, in which
``symmetrization" was thought to play an important role in order
to obtain decoupling and suppression of decoherence
\footnote{Though apparently this point is well appreciated in the
practice of high resolution NMR, i.e., there are many sequences,
e.g., WAHUHA, achieving the intended averaging effect without
averaging over a subgroup. Nevertheless, averaging still results
from symmetry arguments in these cases (L. Viola, private
  communication)}.
Indeed the dynamics (\ref{eq:BBevol}) is different from the
dynamics $[U_1^\dagger U(t/2N) U_1 U(t/2N)]^N$, originally
proposed in \cite{Viola:98}, because it is only constructed with a
single ``bang" $U_1$, without the second ``bang" $U_1^\dagger$
which would close the group. We will further elaborate on this
issue in Sec.\ \ref{sec-cycle}.

By taking $H$ to be a system-bath interaction Hamiltonian, we see
that the effect of the $U_{1}$ ``kicks" is to project the
decohering evolution into disjoint subspaces defined by the
spectral resolution of $U_{1}$. A proper choice of $U_{1}$ can
either eliminate this evolution or make it proceed in some desired
fashion. To give the simplest possible example, suppose
\begin{equation}
H=\sigma _{x}\otimes B, \qquad U_{1}=\sigma _{z}.
\label{eq:bitflip}
\end{equation}
$H$ generates ``bit-flips" and the projection operators are
\begin{equation}
P_{\pm }=\frac{1}{2} (I\pm \sigma _{z})
\end{equation}
with eigenvalues
$\lambda _{\pm }=\pm 1$. Thus
\begin{equation}
H_{Z}=\sum_{\mu =\pm }P_{\mu} HP_{\mu }=
\sum_{\mu =\pm }P_{\mu }\sigma _{x}P_{\mu }\otimes B=0 ,
\end{equation}
so the decohering evolution is completely cancelled.

The physical mechanism giving rise to the Zeno subspaces in the $
N\rightarrow \infty $ limit can be understood by considering the
case of a finite dimensional Hilbert space. Then the limit
(\ref{eq:HZ}) reads
\begin{equation}
\frac{1}{N}\sum_{k=0}^{N-1}U_{1}^{\dagger k}HU_{1}^{k}=\sum_{\mu
,\nu }P_{\mu }HP_{\nu } \frac{1}{N}\sum_{k=0}^{N-1}e^{ik(\lambda
_{\mu }-\lambda _{\nu })}
\label{eq:finitedim}
\end{equation}
and one sees that the last sum is $1$ for $\mu =\nu $ and vanishes
as $O(1/N) $ otherwise. [remember that $\lambda _{\mu }\neq
\lambda_{\nu }\; (\mathrm{mod}\;2\pi )$ for $\mu \neq \nu $ in
Eq.~(\ref{eq:specdec})]. The appearance of the Zeno subspaces is
thus a direct consequence of the fast oscillating phases between
different eigenspaces of the kick. This is equivalent to a
procedure of phase randomization, and is analogous to the case of
strong continuous coupling \cite{Facchi:PRL02}.

\section{Implications of inverse Zeno effect}

The above conclusions are correct in the (mathematical) limit of
large $N$. However it is known that, if $N$ is not too large, the
form factors of the interaction play a primary role and can
provoke an \emph{inverse Zeno effect} (IZE), by which the
decohering evolution is accelerated, rather than suppressed
\cite{IZE,IZEexp}. Reconsider the example (\ref{eq:bitflip}), with
$B$ coupling Q to a generic bath with a thermal spectral density
\begin{equation}
\kappa(\omega) = \int dt \exp (i \omega t) \langle B(t) B
\rangle ,
\label{eq:specdens}
\end{equation}
where $B(t)=e^{i H_B t} B e^{-i H_B t}$ is the interaction-picture
evolved bath operator, $H_B$ the free bath Hamiltonian and
$\langle \ldots \rangle$ the average over the bath state. For
instance, one can consider the linear coupling $B= \int d\omega\;
f(\omega)\left(a(\omega)+ a^\dagger(\omega)\right)$, where
$[a(\omega),a^\dagger(\omega')]=\delta(\omega-\omega')$ are boson
operators and $f(\omega)$ a form factor, while $H_B=\int d\omega\;
\omega a^\dagger(\omega) a(\omega)$. The form factor of the
interaction (together with the bath state) determines the spectral
density (\ref{eq:specdens}). For instance, for an Ohmic bath,
\begin{equation}
\kappa(\omega) \propto
\frac{\omega}{\left(1+\left(\omega/\omega_c\right)^2\right)^n}
\coth \left( \frac{\omega}{2T} \right) ,
\label{eq:ohmic}
\end{equation}
where $\omega_c$ is the frequency cutoff, $T$ the temperature of
the bath (Boltzmann's constant $k=1$) and $n$ an integer $n\geq2$
($n=2$ is typical of quantum dots
\cite{qdots:98}).
The free decay rate is
\begin{equation}\label{fgr}
\gamma = 2\pi \kappa(\omega_0),
\end{equation}
$\omega_0$ being the energy difference between the two qubit
states (Fermi golden rule). The modified decay rate can be shown
to read
\cite{Viola:98,ShiokawaLidar:02}
\begin{eqnarray}\label{eq:gammakickViola}
& &\gamma(\tau)= \lim_{t\to\infty} t
\int_{-\infty}^\infty d \omega \; \kappa(\omega)\;
\nonumber \\
& & \times {\rm sinc}^2 \left(\frac{\omega-\omega_0}{2} t\right)
\tan^2 \left(\frac{\omega-\omega_0}{4} \tau \right)
 \ ,
\end{eqnarray}
where $\tau=t/N$ is the period between kicks and
$\mathrm{sinc}(x)\equiv x^{-1}\sin x$. By expanding for large
values of $N$ one gets \cite{deco}
\begin{eqnarray}
 \gamma(\tau)
 \sim \frac{8}{\pi}\,
\kappa\left(\frac{2\pi}{\tau}\right) , \qquad \tau \to 0 \ .
 \label{eq:gammaklimexp}
\end{eqnarray}
Notice that, according to (\ref{eq:gammaklimexp}), for small
values of $\tau$ the modified decay rate $\gamma(\tau)$ is
proportional to the ``tail" of the spectral density
$\kappa(\omega)$. By defining a characteristic \emph{transition}
time $\tau^*$, solution of the equation
\begin{equation}\label{eq:taujumpexp}
\kappa\left(\frac{2\pi}{\tau^*}\right) \simeq \frac{\pi}{8}\gamma
= \frac{\pi^2}{4} \kappa(\omega_0) ,
\end{equation}
one obtains
\begin{eqnarray}
& &\gamma(\tau)<\gamma \qquad \mbox{for}
\quad \tau<\tau^* ,
\nonumber\\
& &\gamma(\tau)>\gamma \qquad \mbox{for}
\quad \tau>\tau^* .
\end{eqnarray}
Decoherence is suppressed in the former case, but it is
\emph{enhanced} in the latter situation (which is analogous to
what one calls IZE in the case of projective measurements). This
shows that an ``inverse Zeno regime" is a serious drawback also in
the case of dynamical decoupling. Since the limit $\tau<\tau^*$
can be very difficult to attain, for a
\emph{bona fide} dissipative system, the efficacy of BB as a
method for decoherence suppression must be carefully analyzed. For
instance, in the Ohmic case (\ref{eq:ohmic}) at low temperature $T
\ll
\omega_0 \ll
\omega_c$, one easily gets
from (\ref{eq:gammaklimexp})
\begin{equation}
\tau^*\simeq 2\pi\omega_c^{-1}
\left(\frac{\pi^2}{4}\frac{\omega_0}{\omega_c}\right)^{\frac{1}{2n-1}}
 \ll 2\pi \omega_c^{-1},
\label{eq:izlimit}
\end{equation}
a condition that may be difficult to achieve in practice. In fact,
we see here that the relevant timescale is not simply the inverse
bandwidth $\omega_c^{-1}$, but can be much shorter if $\omega_0
\ll \omega_c$, as is typically the case. It has already been
observed that the Ohmic bath is a particularly demanding setting
for BB, and that spin-boson baths with decaying spectral density
$I(\omega)$ [not to be confused with the thermal spectral density
$\kappa(\omega)$], such as $1/f$, are more amenable to successful
BB decoupling \cite{ShiokawaLidar:02}. We will reconsider this
issue from the point of view of the IZE in \cite{deco}.

\section{BB cycle of several pulses}
\label{sec-cycle}

We now generalize the previous result to the situation where each
cycle consists of $g$ kicks. This will allow us to show how the
procedure of ``decoupling by symmetrization'' \cite{Viola:99},
i.e., the standard view of the BB effect, arises as a special case
of such cycles and is related to the QZE. We consider $N$ cycles
of $g$ instantaneous kicks $U_{1},\dots ,U_{g}$ in a time interval
$t$
\begin{equation}
U_{N}(t)=\left[ U_{g}U\left( \frac{t}{gN}\right) \cdots U_{2}U\left( \frac{t
}{gN}\right) U_{1}U\left( \frac{t}{gN}\right) \right] ^{N}.
\label{eq:kickseq}
\end{equation}
We use the same notation as above, sometimes abbreviating
$U(t/gN)$ by $U$, unless confusion may arise. Similarly to the
single-kick case, in the $N\rightarrow \infty $ limit, the
dominant contribution is $(U_{g}\cdots U_{2}U_{1})^{N}$ and it is
convenient to consider the sequence of unitary operators
\begin{equation}
V_{N}(t)=(U_{g}\cdots U_{1})^{\dagger N}U_{N}(t).
\end{equation}
The differential equation is again
\begin{equation}
i\frac{d}{dt}V_{N}(t)=H_{N}(t)V_{N}(t),\qquad (V_{N}(0)=I)
\end{equation}
where
\begin{eqnarray}
H_{N}(t)& = & \frac{1}{N}\sum_{k=0}^{N-1}(U_{g}\cdots
U_{1})^{\dagger
N}(U_{g}U\cdots U_{1}U)^{k} \nonumber \\
& & \times \bar{H}_{N}(U_{g}U\cdots U_{1}U)^{\dagger
k}(U_{g}\cdots U_{1})^{N},  \label{eq:HNt}
\end{eqnarray}
with
\begin{eqnarray}
& & \bar{H}_{N} =\frac{1}{g}\left[ U_{g}HU_{g}^{\dagger}+
(U_{g}UU_{g-1})H(U_{g}UU_{g-1})^{\dagger }+\cdots \right.
\nonumber \\
& & \left.  +(U_{g}UU_{g-1}\cdots
U_{2}UU_{1})H(U_{g}UU_{g-1}\cdots U_{2}UU_{1})^{\dagger }\right] .
\end{eqnarray}
We can now follow through the same calculations as in the
single-kick case, substituting $U_{1}HU_{1}^{\dagger }$ everywhere
by $\bar{H}_{N}$, and $U_{1} $ by $U_{g}\cdots U_{1}$. It is then
straightforward to verify that in the $ N\rightarrow \infty $
limit we get
\begin{equation}
\mathcal{U}(t)\equiv \lim_{N\rightarrow \infty }V_{N}(t),
\end{equation}
which again satisfies Eq.~(\ref{eq:eqUz}), with the Zeno
Hamiltonian
\begin{equation}
H_{Z}=\Pi _{U_{g}\cdots U_{1}}(\bar{H})=\sum_{\mu }P_{\mu
}\bar{H}P_{\mu },
\label{eq:HZZ}
\end{equation}
where
\begin{eqnarray}
& & U_{g}\cdots U_{2}U_{1}=\sum_{\mu }P_{\mu }e^{-i\lambda _{\mu
}},
\label{eq:Ug} \\
& & \bar{H}=
\frac{1}{g}[H+\cdots +(U_{g-2}\cdots U_{1})^{\dagger
}H(U_{g-2}\cdots U_{1}) \nonumber \\
& & +(U_{g-1}\cdots U_{1})^{\dagger }H(U_{g-1}\cdots U_{1})].
\label{eq:Hbar}
\end{eqnarray}
In conclusion,
\begin{eqnarray}
U_{N}(t) &\sim & (U_{g}\cdots
U_{1})^{N}\mathcal{U}(t)=(U_{g}\cdots U_{1})^{N}\exp (-iH_{Z}t)
\nonumber \\
&=& \exp \left( -i\sum_{\mu }(N\lambda _{\mu }P_{\mu }+P_{\mu
}\bar{H}P_{\mu }t)\right) .
\end{eqnarray}
It is clear that also in this case we get a QZE, with relevant
Zeno subspaces \cite{Facchi:PRL02}. The only difference from the
single-kick case is that the Hamiltonian $\bar{H}$
[Eq.~(\ref{eq:Hbar})] and the product of the cycle $U_g\cdots U_2
U_1$ [Eq.~(\ref{eq:Ug})] take the place of $H$ and $U_1$,
respectively.

It is important to observe again that \emph{no symmetry} or group
structure is required from the ``kick'' sequence
(\ref{eq:kickseq}): the above formulas are of general validity, as
they rely on the von Neumann ergodic theorem. They reduce to the
usual expression in the case of a finite closed group of unitaries
$\mathcal{G}$ with elements $ V_{r}$, $r=1,\dots ,g$ and
$V_{1}=I$. Indeed, decoupling by symmetrization \cite{Viola:99} is
recovered as a particular case by considering the unitary
operators
\begin{equation}
U_{r}=V_{r+1}V_{r}^{\dagger },\quad (r=1,\dots ,g-1),\qquad
U_{g}=V_{g}^{\dagger }.
\end{equation}
A single cycle yields
\begin{equation}
U_{\mathrm{cycle}}(t)=V_{g}^{\dagger }U\left( \frac{t}{gN}\right)
V_{g}\cdots V_{1}^{\dagger }U\left( \frac{t}{gN}\right) V_{1},
\end{equation}
while
\begin{equation}
U_{g}\cdots U_{1}=V_{g}^{\dagger }V_{g}V_{g-1}^{\dagger }\cdots
V_{2}^{\dagger }V_{2}=I.
\end{equation}
We therefore reobtain, as a special case of the QZE, the
well-known BB result \cite{Viola:99}:
\begin{equation}
U_{N}(t)=V_{N}(t)\overset{N\rightarrow \infty }{\sim }\exp
(-iH_{\mathrm{eff} }t),
\end{equation}
where $H_{\mathrm{eff}}=H_{Z}$ and
\begin{equation}
H_{Z}=\Pi _{I}(\bar{H})=\bar{H}=\frac{1}{g}
\sum_{r=1}^{g}V_{r}^{\dagger }HV_{r}=\Pi _{\mathcal{G}}(H).
\end{equation}

\section{Origin of equivalence between continuous and
pulsed formulations}

The equivalence between the ways in which the QZE can be generated
via observation and via Hamiltonian interaction have been
discussed in \cite{Facchi:PRL02}. We now explain the equivalence
between the continuous and pulsed Hamiltonian interaction
pictures, in generating the Zeno subspaces. In fact, the two
procedures differ only in the order in which two limits are
computed. We recall that the continuous case deals with the strong
coupling limit \cite{Facchi:PRL02}
\begin{equation}
H_{\mathrm{tot}}=H+KH_{1},\qquad K\rightarrow \infty   \label{eq:cont}
\end{equation}
and the Zeno subspaces are the eigenspaces of $H_{1}$. On the
other hand, the kicked dynamics entails the limit $N\rightarrow
\infty $ in (\ref{eq:BBevol}) and the Zeno subspaces are the
eigenspaces of $U_{1}$. This evolution is generated by the
Hamiltonian
\begin{equation}
H_{\mathrm{tot}}=H+\tau _{1}H_{1}\sum_{n}\delta (t-n\tau _{2}),\qquad \tau
_{2}\rightarrow 0  \label{eq:puls}
\end{equation}
where $\tau _{2}$ is the period between two kicks and the unitary
evolution during a kick is $U_{1}=\exp (-i\tau _{1}H_{1})$. The
limit $N\rightarrow \infty $ in (\ref{eq:BBevol}) corresponds to
$\tau _{2}\rightarrow 0$. The two dynamics (\ref{eq:cont}) and
(\ref{eq:puls}) are both limiting cases of the following one
\begin{eqnarray}
  H_{\mathrm{tot}} &=& H+KH_{1}
  \sum_{n}g\left( \frac{t-n(\tau _{2}+\tau _{1}/K)}{\tau
  _{1}/K}\right) ,
\label{eq:contpuls}
\end{eqnarray}
where the function $g$ has the properties
\begin{eqnarray}
\sum_{n}g(x-n) &=& 1 \label{eq:prop1}\\
\lim_{K\rightarrow \infty} K g(K x) &=& \delta (x).
  \label{eq:prop2}
\end{eqnarray}
For example we can consider $g(x)=\chi_{[-1/2,1/2]}(x)$, where
$\chi_{I}$ is the characteristic function of the set $I$. In Eq.\
(\ref{eq:contpuls}) the period between two kicks is $\tau
_{1}/K+\tau_{2}$, while the kick lasts for a time $\tau _{1}/K$.
By taking the limit $\tau _{2}\rightarrow 0$ in Eq.\
(\ref{eq:contpuls}), i.e., a sequence of pulses of finite duration
$\tau_1/K$ without any idle time among them, and using property
(\ref{eq:prop1}), one recovers the continuous case
(\ref{eq:cont}). Then, by taking the strong coupling limit
$K\rightarrow \infty $ one gets the Zeno subspaces. On the other
hand, by taking the $K\rightarrow \infty $ limit, i.e., the limit
of shorter pulses (but with the same global---integral---effect),
and using property (\ref{eq:prop2}) and the identity $\delta
(t/\tau_{1})=\tau _{1}\delta (t)$, one obtains the kicked case
(\ref{eq:puls}). Then, by taking the vanishing idle time limit
$\tau _{2}\rightarrow 0$ one gets again the Zeno subspaces. In
short, the mathematical equivalence between the two approaches is
expressed by the relation
\begin{equation}
\label{eq:limits}
  \lim_{K\rightarrow \infty }\;\lim_{\tau _{2}\rightarrow
    0}H_{\mathrm{tot}}
  =\lim_{\tau _{2}\rightarrow 0}\;\lim_{K\rightarrow \infty }H_{\mathrm{tot}},
\end{equation}
(for almost all $\tau_1$) with the left (right) side expressing
the continuous (pulsed) case. Note that this formal equivalence
must physically be checked on a case by case basis, and it is
legitimate only if the inverse Zeno regime is avoided and the role
of the form factors clearly spelled out. That is, physically the
relevant timescales play a crucial role, and in practice there
certainly can be a difference between kicked dynamics and
continuous coupling, in spite of their equivalence in the above
mathematical limit.

Another key issue of physical relevance, in particular if one is
interested in possible applications, is played by the physical
meaning of ``strong'' when one talks of the strong coupling
regime. We showed that strong coupling is equivalent to large $N$
(number of interruptions) and, since experiments with large $N$
have been performed, proving both the quantum Zeno and the inverse
quantum Zeno effect \cite{IZEexp}, the strong coupling regime is
attainable in real physical systems.

\section{Conclusions}

In this work we have shown the formal equivalence of the quantum
Zeno effect (QZE), which has been known since von Neumann laid
down the mathematical foundations of quantum mechanics
\cite[p.366]{vonNeumann:55} and has been the subject of intense
investigations since the seminal paper \cite{Misra:77}, to the
recently introduced \cite{Viola:98,Vitali:99,Viola:99}
``bang-bang'' decoupling method (BB) for reducing decoherence in
quantum information processing \footnote{In fact the original BB
paper \cite{Viola:98} recognized the mathematical connection to
the QZE, in particular the features of Cook's method for the
inhibition of a stimulated two-level transition by pulsed
measurements
\cite{Cook}, but stated that ``the analogy stops from a more
physical point of view''. }. The QZE is traditionally derived by
considering a series of rapid, \emph{pulsed observations}
\cite{Misra:77}. This became almost a dogma and motivated
interesting seminal experiments \cite{Cook,IZEexp}. Later
formulations emphasized that the QZE can also be generated by
\emph{continuous Hamiltonian interaction}
\cite{Harris:82plSchulman98,Facchi:PRL02,Beige:00}. The BB method,
on the other hand, employs a series of rapid \emph{pulsed
interactions}. Here we have shown that both the QZE (in its
continuous-interaction formulation) and the BB method can be
understood as limits of a single Hamiltonian,
Eq.~(\ref{eq:contpuls}), giving rise to either pulsed or
continuous dynamics, with a resulting partitioning of the
controlled system's Hilbert space into quantum Zeno subspaces,
defined by Eqs.~(\ref{eq:specdec})-(\ref{eq:HZ}). This unified
view not only offers the advantage of conceptual simplicity, but
also has significant practical consequences: it shows that the
scope of all the methods analyzed here (QZE, BB and continuous
interaction) are wider than previously suspected, leading to
greater flexibility in their implementation. In particular, since
all formulations of the QZE are physically equivalent, and BB is
equivalent to the kicked unitary formulation of the QZE, it is
clear that BB can also be formulated in terms of a
\emph{continuous interaction} and
\emph{pulsed measurements}. The continuous interaction version of
BB avoids the frequently criticised off-resonant transitions
associated with the large bandwidth pulses required in the pulsed
BB implementation \cite{Tian:00}. We have not studied the
practical advantages or drawbacks of the  pulsed measurement
formulation of BB.

We emphasize that our conclusions about greater flexibility in the
practical implementation of the BB method are supported by the
fact that experiments with large $N$ have been performed, proving
both the quantum Zeno \cite{IZEexp,Cook} and the inverse quantum
Zeno effect \cite{IZEexp}, and showing that the strong coupling
regime is attainable in real physical systems.

Another consequence of our work is that the Zeno-subspace
dynamics, in its pulsed formulation, can be generated by a
sequence of \emph{arbitrary} (fast and strong) pulses, without any
(symmetry) assumptions about the relation between pulses. This
generalizes all previously published formulations of the BB
method, which assumed such relations.

Finally, owing perhaps to its longer history, the QZE has been
more thoroughly studied than the BB method, and it has been
recognized that in physically relevant limits an inverse QZE can
arise. We have shown that the same conclusion applies to the BB
method, with the important implication that in some cases BB can
actually enhance, rather than reduce decoherence. This issue will
be the subject of further investigations \cite{deco}.

\begin{acknowledgments}
We thank the organizers of the Conference on ``Irreversible
Quantum Dynamics,'' (August 2002) who brought us together in
Trieste. D.A.L. is supported in part by the DARPA-QuIST program
(managed by AFOSR under Grant No. F49620-01-1-0468).
\end{acknowledgments}

\end{document}